\documentclass[aps,prl,10pt,twocolumn,floatfix,amsmath,amssymb,longbibliography,raggedbottom]{revtex4-2}
\usepackage{graphicx}
\usepackage{amsmath,amssymb}
\usepackage{bm}

\begin{document}

\title{Fundamental constraints on the observability\\ of non-Hermitian  effects in passive systems}

\author{Henning Schomerus}
\affiliation{Department of Physics, Lancaster University, Lancaster, LA1 4YB, United Kingdom}

\begin{abstract}
Utilizing scattering theory, we
quantify the consequences of  physical constraints that limit the visibility of non-Hermitian effects in passive devices. The constraints arise from the fundamental requirement that the system obeys causality, and can be captured concisely in terms of an internal time-delay operator, which furthermore provides a direct quantitative measure of the visibility of  specific non-Hermitian phenomena in the density of states. We
illustrate the implications by contrasting different symmetry classes
and non-Hermitian effects, including exceptional points and the non-Hermitian skin effect, whose underlying extreme mode nonorthogonality turns out to be effectively disguised.
\end{abstract}

\maketitle

Hermitian systems support orthogonal stationary modes with identical, infinite, life times. If the Hermiticity is of a fundamental nature, as in quantum mechanics, this plays an important role in guaranteeing the dynamical stability of the system. The situation is more complex in effectively non-Hermitian systems \cite{moiseyev2011non,ashida2020}, for instance photonic systems with gain or loss \cite{Cao15,Feng2017,ElG18,Midya2018,Ota2020}, where nonorthogonal modes of different life times appear.
These systems attract considerable attention because they can display a wide range of  phenomena that can be exploited for unique applications, such as power oscillations \cite{makris2008,Guo2009,Rueter2010}, unidirectional transport \cite{Ram10} and invisibility \cite{Lin2011}, coherent absorption \cite{Cho10,Longhi10},  mode selection \cite{Pol15} and lasing \cite{Lie12,Feng2014,Hodaei2014,Peng14,Zha18,Par18}. Many of these applications purposefully utilize the life-time differences to enhance desired modes relative to undesired modes---hence, achieve a clear visibility of specific modes  incorporating a particular functionality by their long life time. Furthermore, many of these applications make explicit use of the mode nonorthogonality. Prominent examples are enhanced sensors \cite{Wie14,Che17,Wiersig20} operating near exceptional points \cite{Berry2004,miri2019,oezdemir2019}, which are non-Hermitian degeneracies where eigenmodes align, and directed amplifiers and sensors \cite{Sch20,Budich2020,Wanjura2020,McDonald2020,Midya2022} facilitated by the non-Hermitian skin effect \cite{hatano1996,Yao18,Kun18,Lee19,Borgnia2020},  a prominent feature of non-reciprocal systems where all bulk modes become systematically distorted towards one side of the system.

The design of such systems has received further impetus by the realization that in non-Hermitian systems, modes with predetermined frequencies, life times, and mode profiles can be protected by generalised symmetries  \cite{Sch13,Lieu2018,Kaw19,Oza19}. This leads to rich scenarios that transcend Hermitian topological physics \cite{Has10,Qi11,Beenakker2015}, both practically \cite{Price2022} as well as in their mathematical complexity  \cite{Gon18,Ber19,Okuma2022}.
However, many of these symmetries can only be realized in active systems, which require a sustained supply of energy and introduce noise.
For instance, parity-time (PT) symmetric systems with balanced gain and loss \cite{Feng2017,ElG18} can  provide a spectrum of infinite life-time modes, but these are intrinsically destabilised by the quantum noise in the active regions \cite{Sch10,Yoo11,Scheel2018}.
Furthermore, present  experimental realizations of the non-Hermitian skin effect \cite{Bra19,ghatak2020,Helbig2020,Weidemann2020,liang2022observation} all invoke active elements. Analogously, noise limits the precision of exceptional-point sensors \cite{Langbein2018}, and non-adiabatic transients from life-time differences limit the observability of the half-integer Berry phase of these points \cite{Uzdin2011,Doppler2016}. Even in the passive setting, many of the most coveted non-Hermitian effects can be made visible only with specially tailored excitations.
In view of these challenges, it is highly desirable to base non-Hermitian functionality on the generic scattering response of passive stationary devices in which specific modes display long life times, both absolutely as well as relatively when compared to other modes.

In this work, we
establish and evaluate a stringent fundamental constraint on this objective, which arises from the requirement that the underlying microscopic physics obeys causality  \cite{Toll1956}. This requirement is stronger than just insisting on the dynamical stability of the modes in the system (hence, on non-negative lifetimes), and takes care of the fact that the system is not isolated, but couples to the components supplying the loss. However, the constraints can be readily formulated in general terms, and evaluated quantitatively in given systems. From this we can determine general limits of the visibility of specific non-Hermitian effects.

To establish these links systematically, we first formulate the constraints compactly by phrasing them in terms of the language of scattering theory. Causality is then encoded into the internal time-delay operator, a central object from scattering theory that can be obtained from the microscopic model. From this, one can obtain the critical threshold value of overall  losses that a model necessarily needs to include to be realizable in a passive device.
These conditions can be classified by symmetries inherited from the effective Hamiltonian, which establishes a systematic link between non-Hermitian and Hermitian symmetry classes.
Furthermore, the time-delay operator determines a spectral function (the scattering density of states) that serves as a quantitative measure of the visibility of specific non-Hermitian effects. We
apply this to scenarios of particular theoretical and experimental interest,
comprising systems displaying exceptional points, the non-Hermitian skin effect, and edge states.
Remarkably, in the passive setting, the extreme mode nonorthogonality underlying exceptional points and the skin effect turns out to be effectively disguised.

\emph{General formulation of causality constraints.}
For concreteness,
we base the considerations on coupled-mode theory, which enjoys a wide range of applications from photonic to mechanical systems and mimics the language of quantum mechanics. In this theory, one employs a basis corresponding to a suitable set of bare modes, often identified with individual  components such as resonators or waveguides, and collects their intrinsic properties and couplings in a matrix $H$ that serves as an effective Hamiltonian. We assume that the elements of $H$ are frequency-independent, but note that effects of additional frequency dispersion can be accounted for by including auxiliary components \cite{Keil2016}.
This effective Hamiltonian serves as the input to model and design experiments, and also is the ubiquitous starting points for theoretical considerations, such as about the role of symmetries in these systems, as specified further below. For the moment,  the key feature of the effective Hamiltonian is that it can be non-Hermitian, $H\neq H^\dagger$. On the basis of the microscopic model, we capture this quantitatively by separating out the Hermitian and nontrivial anti-Hermitian parts as well as an overall level of uniform scalar background losses, $H=H_0+iF-i\gamma$, where $H_0=H_0^\dagger$,
$F=F^\dagger$, and $\gamma$ set so that $\mathrm{tr}\,F=0$.
As we will see, the model becomes realizable in a passive device when $\gamma$ exceeds a certain threshold value $\gamma_c$, which we will call the \emph{causality threshold}, and constitutes a central quantity in our considerations.

To identify this threshold and establish the ensuing limits on the visibility of non-Hermitian effects, we probe the system from the outside and adopt a scattering perspective.
Let us assume that the system is uniformly coupled to the outside at a coupling strength $\Gamma$, and state as our aim to resolve the desired modes in the scattered signal. In the wide-band limit, where all spectral features in the scattered signal are due to the system, the standard scattering matrix is given by
\begin{align}
S(\omega) &=(1-i\Gamma \mathcal{G}(\omega))(1+i\Gamma \mathcal{G}(\omega))^{-1}
,
\label{eq:s}
\end{align}
where
\begin{align}
\mathcal{G}(\omega)&=(\omega-H_0-iF+i\gamma)^{-1}
\label{eq:g}
\end{align}
is the Greens function of the system.
From the scattering perspective, the system is passively realizable if it does not amplify any incoming signal $\mathbf{a}$, i.e., if $||S\mathbf{a}||\leq ||\mathbf{a}||$, irrespective of whether the incoming signal is designed to couple into a specific mode or not. This means that the operator combination $S^\dagger S$ has no eigenvalues exceeding 1, or equivalently, that the expression
$1-S^\dagger S$
is positive semidefinite. With Eq.~\eqref{eq:s}, this condition can be reformulated instructively by writing
\begin{equation}
1-S^\dagger S=2\Gamma Q_\Gamma,
\label{eq:sq}
\end{equation}
where
\begin{equation}
Q_\Gamma=2\mathcal{G}(\omega+i\Gamma)^\dagger(\gamma-F)\mathcal{G}(\omega+i\Gamma)
\label{eq:q}
\end{equation}
is the time-delay operator in an open system \cite{Beenakker2001}. Via Eq.~\eqref{eq:sq}, its elements can be directly determined in experiments from the scattering strength.
For a passively realizable system we then have to demand that $Q_\Gamma$ is positive semidefinite.

This formulation is useful because it provides a unifying perspective on different aspects of the system.

Firstly, it confirms that on the simplest level, a microscopic model can be physically realized in a passive system when the combination $\gamma-F$ itself is positive semidefinite \cite{Wiersig2019}, hence, if $\gamma$ is larger than the largest eigenvalue of $F$. We denote the eigenvalues of $F$ as $f_k$, so that $\gamma_c=\mathrm{max}\,f_k$. This constraint is both \emph{simpler} and \emph{stronger} than requiring positive life times $\tau_k$ of all modes, which are encoded in the eigenvalues $\Omega_k=\omega_k-i/2\tau_k$ of the effective Hamiltonian $H$ itself, and related mathematical constraints such as the Lee-Wolfenstein and Bell-Steinberger relations \cite{Wiersig2019,Lee1965,Bell1966}.

Secondly, the positive semidefiniteness of the delay times guarantees causality of the system, which clarifies the fundamental physical nature of the constraint, and justifies to call $\gamma_c$ the  causality threshold.

Thirdly, the time-delay operator delivers a direct measure of the generic visibility of physical effects, the scattering density of states $\rho(\omega)=(2\pi)^{-1}\,\mathrm{tr}\,Q_\Gamma(\omega)$, which accounts for the mode broadening by the intrinsic and radiative losses.
We read off that for fixed $H_0$ and $F$, the passive visibility of individual modes in this measure is maximized for $\Gamma=0$ and $\gamma=\gamma_c$, hence at the causality threshold of a weakly probed system.
Below, we will use this measure to quantify the properties of specific systems from different symmetry classes, whose general features we describe next.

\emph{Evaluation in specific symmetry classes.}
Effectively non-Hermitian Hamiltonians appear in many guises, and the physical symmetries of a system translate into different mathematical forms that reflect the context.
E.g., the effective Hamiltonian may generate the time evolution of a wavefunction in a slowly-varying envelope coupled-mode description, which is analogous to the Schr{\"o}dinger equation, it may feature in the propagation of a density matrix, or it may appear in a stability analysis or Bogoliubov theory that includes complex-conjugated fields \cite{Lieu2018,Kaw19}. In the identification of the causality constraints above, we adopted the language of scattering theory, and we therefore apply the corresponding notions when identifying the symmetries of the system \cite{Pikulin2013,schomerus2013from,Beenakker2015}.
Specifically, conventional time-reversal (T) symmetry then entails $H=H^*$, and in a reciprocal system we have $H=H^T$ (we will refer to this as the T$'$ symmetry). Furthermore, PT symmetry dictates $\mathcal{X}H\mathcal{X}=H^*$, and a non-Hermitian charge-conjugation (C) symmetry entails  $\mathcal{X}H\mathcal{X}=-H^*$, involving in each case a suitable unitary operator fulfilling $\mathcal{X}^2=1$. We will also consider variants of these systems with $\mathcal{X}H\mathcal{X}=H^\dagger$ (PTT$'$) and $\mathcal{X}H\mathcal{X}=-H^\dagger$ (CT$'$), which constitute separate symmetry classes when the system is non-reciprocal.

It should be noted that several of these symmetries can only be  realized for $\gamma=0$. To extend them to the passive setting, we hence assume that they hold for $H=H_0+iF$, and then consider the same model at finite $\gamma$, following the established example of passive PT symmetry \cite{Guo2009}. The symmetry can then be realized passively for $\gamma\geq\gamma_c$.
These symmetries can be combined in different ways, leading to an extensive classification \cite{Lieu2018,Kaw19}, where the cases above are of particular theoretical and experimental interest. For instance, non-reciprocal systems with $H=H^*\neq H^T$ are the simplest setting in which the non-Hermitian skin effect appears.

\begin{figure}
  \includegraphics[width=\linewidth]{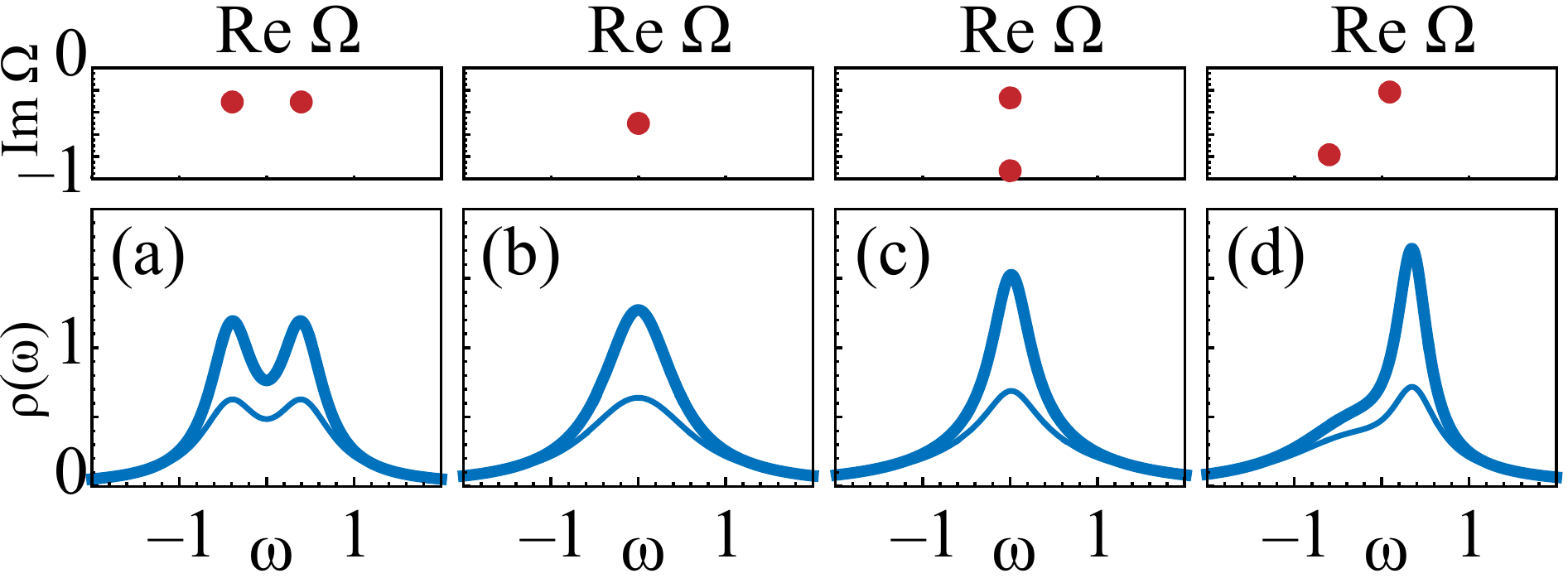}
  \caption{Disguised exceptional points in passive non-Hermitian systems. Complex eigenvalue spectrum $\Omega_k$ (top) and density of states $\rho(\omega)$ (bottom) for the effective Hamiltonian \eqref{eq:HEP}. The couplings are fixed to $b=c^*=0.4-0.3i$, while (a) $a=\sqrt{-bc}-0.2 i$ (PT symmetric case),
  (b) $a=\sqrt{-bc}$ (EP),  (c) $a=\sqrt{-bc}+0.1 i$ (dynamically broken phase), and (d) $a=\sqrt{-bc}+0.2$ (symmetry explicitly broken).
  The thick lines are for the limit of weak couplings $\Gamma=0$, where the life times are maximized. At the exceptional point, $\rho(\omega)$ is then a simple Lorentzian, which disguises the extreme mode nonorthogonality of the system.
The thin lines are for finite $\Gamma=0.1$, where a squared-Lorentzian background of limited contrast appears at the EP.
  }\label{fig:EP}
\end{figure}

Before we address the visibility of such effects, let us examine the general structure of the causality constraints in these symmetry classes. We express this compactly in terms of the symmetries inherited by the operator $F$ capturing the nontrivial non-Hermitian content of the model, and refer to \footnote{
See the Appendix 
for concrete constructions of the operator $F$ in different symmetry classes
 as well as additional analytical and numerical results.}
 for a detailed description in terms of the block structure of this operator.

We start with the case of a PT-symmetric system. From the definition of the symmetry in terms of the effective Hamiltonian, we see that the operator $F$ then obeys a Hermitian charge-conjugation symmetry,
$\mathcal{X}F\mathcal{X}=-F^*$, as encountered in a superconductor \cite{Beenakker2015}. This enforces a symmetry of its spectrum, with eigenvalues appearing in pairs $\pm |f_k|$.
For a PTT$'$ symmetry,
the operator $F$
displays a chiral symmetry,  $\mathcal{X}F\mathcal{X}=-F$, again leading to a spectral symmetry of its eigenvalues.
Both variants coincide if the system is reciprocal, $H=H^T$, where $F$ is real and obeys time-reversal symmetry.

For systems with a C symmetry, $
F  =\mathcal{X}F^*\mathcal{X}
$
displays a generalized  time-reversal symmetry,
while CT$'$ entails that
$F=\mathcal{X}F\mathcal{X}$ obeys a unitary symmetry, and hence
can be block-diagonalized into the symmetry sectors of $\mathcal{X}$.
Finally, in a non-reciprocal system with passive $T$ symmetry,
$
F=-F^T$, as is typical for a topologically nontrivial superconductor in the so-called Majorana basis \cite{Beenakker2015}.
In all cases, we see that there is a systematic link from a non-Hermitian symmetry class to a Hermitian symmetry class.

This concludes or first main objective of formulating a convenient general framework.
We now apply it to quantify the visibility of non-Hermitian effects in concrete settings.

\emph{Observability of specific non-Hermitian effects.}
As mentioned in the introduction, exceptional points (from here on EP's), where eigenvalues collide in the complex plane and eigenmodes align, are one of the most prominent features of non-Hermitian systems.
Their effects are most easily seen for modes of long life times, such as in active photonic systems near the laser threshold \cite{Lie12,Feng2014,Hodaei2014,Peng14}, which then are highly sensitive to perturbations \cite{Wie14,Che17,Wiersig20}.
The enhanced sensitivity also applies to the noise from spontaneous emission, which at the EP results in an unconventional \emph{squared-Lorentzian} lineshape \cite{Yoo11,Takata21}.
The squared Lorentzian also shows up as a background when a passive system is forced into a stationary state by external driving \cite{Hashemi2022,Wiersig2022}.
This behaviour is linked to the drastic violation of mode orthogonality at the EP.

\begin{figure*}
  \includegraphics[width=\textwidth]{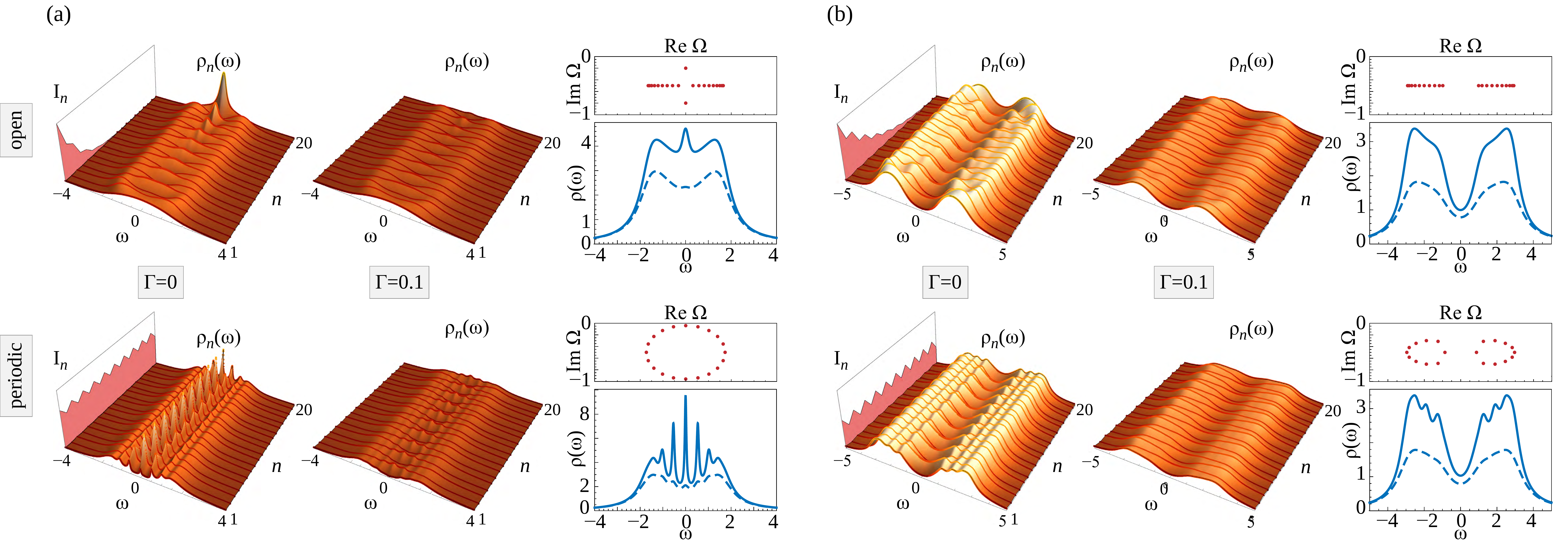}
  \caption{Disguised non-Hermitian skin effect, illustrated by model
\eqref{eq:Hnm} in finite systems of $20$ sites, $v_1=-0.8i$, $v_2=-0.2i$, $u_1^\pm=\bar u\mp 0.4$, $u_2^\pm=1$, $w=0$, and (a) $\bar u=0.8$, (b) $\bar u=2$, with open (top) or periodic (bottom) boundary conditions.
Here $\rho_{n}(\omega)$ and $\rho(\omega)$ are the spatially resolved and overall density of states, for which we contrast the case $\Gamma=0$  (limit of vanishing coupling, left supanels/solid curves) and $\Gamma=0.1$ (strong coupling, right supanels/dashed curves). This physical data is compared to the
summed eigenvector profile $I_n$  displaying the skin effect, and the complex eigenvalues $\Omega$  of $H$.
 }\label{fig:NHSE}
\end{figure*}

To assess the observability of these features in the scattering density of states, we adopt the standard reduced $2\times 2$ Hamiltonian
\begin{equation}
H=\begin{pmatrix}
        a-i \gamma& b \\
        c & -a -i \gamma
\end{pmatrix},
\label{eq:HEP}
\end{equation}
where we initially allow for the most general case with complex parameters $a$, $b$, and $c$.
The causality constraint then follows from the matrix
\begin{equation}
F=(-i/2)\begin{pmatrix}
        a-a^*& b-c^* \\
        c-b^* & -a+a^*
\end{pmatrix},
\end{equation}
giving the threshold value $\gamma_c=\sqrt{(\mathrm{Im}\,a)^2 +|b-c^*|^2/4}$.

In Fig.~\ref{fig:EP} we show the threshold scattering density of states $\rho(\omega)$ for different scenarios at and away from the EP, which occurs for $a^2+bc=0$.  In the figure, the EP signals a spontaneous symmetry-breaking transition, where the eigenvalue  pair  moves away from the spectral symmetry line $\mathrm{Im}\,\omega=-\gamma$.
At the EP, this density of states is given by
\begin{equation}
\rho^{(\mathrm{EP})}(\omega)=\frac{1}{\pi}\frac{|b|+|c|}{\omega^2+(|b|+|c|)^2/4}.
\label{eq:rhoep}
\end{equation}
This is a \emph{simple} Lorentzian normalized to 2, hence accounts for both states but fully disguises the extreme mode nonorthogonality at the degeneracy.
A  squared-Lorentzian background of limited contrast only appears at finite coupling $\Gamma$ (dashed lines). However, this background comes with a negative spectral weight that never exceeds 1/4 of the Lorentz contribution \cite{Note1}, with the optimal contrast attained at $\Gamma\sim\gamma_c$ where the density of states is already strongly suppressed due to the much reduced lifetime to the states.

In other symmetry classes, functionality can arise collectively from the bulk, or individually from particular states. To determine their visibility we consider a  flexible one-dimensional model encompassing a wide range of paradigms \cite{Esa11,Ramezani2012,Sch13,Longhi2015,Mal15,Lee2016,Yao18,Ni2018,Mostafavi2020,Ghaemi21}, based on an effective Hamiltonian
\begin{align}
H_{nm}&=\delta_{n,m}v_n+\delta_{n,m-1}u_n^-+\delta_{n,m+1}u_m^+
\nonumber\\
&{}+\delta_{n,m-2}w_n+\delta_{n-2,m}w_m.
\label{eq:Hnm}
\end{align}
The imaginary parts of the complex onsite-potentials $v_n$ induce distributed losses, while the real parts allow to define a corrugated potential. The couplings  $u_n^\pm$ can be arranged to induce zero modes, and when they are non-reciprocal they induce the non-Hermitian skin effect. For the moment, we also include reciprocal real next-nearest neighbour couplings $w_n$ as they allow to obtain the skin effect from scalar losses and real magnetic fields \cite{Longhi2015}, and feature in some experiments \cite{Bra19}.

All these effects appear in a periodic dimer arrangement with a two-site unit cell,
which induces a non-trivial band structure governed by the Bloch Hamiltonian
\begin{equation}
H(k)=\begin{pmatrix}
        v_1+2w_1\cos k& u_1^- + u_2^+ e^{-ik} \\
        u_1^+ + u_2^- e^{ik} & v_2+2w_2\cos k
\end{pmatrix}.
\end{equation}
With conventional periodic boundary conditions, $k$ is real, but
to fulfill the open boundary conditions of a finite system one generically needs to combine spectrally degenerate non-reciprocal modes with complex $k$ \cite{Imu19,Yokomizo19,Yang2020non}. Hence, the modes display an exponential spatial profile distorted toward the edge, which is the essence of the non-Hermitian skin effect. In the Hermitian limit, these modes become conventional extended Bloch states, with the possible exception of a finite number of edge states that maintain an exponentially decaying profile.
This complex interplay of effects explains why the non-Hermitian skin effect has received considerable attention.

In contrast, the causality constraints on the bulk and edge modes are governed much more simply by the Bloch version of the operator $F$,
\begin{equation}
F(k)=\begin{pmatrix}
        \Delta v& \Delta u_1^* + \Delta u_2 e^{-ik} \\
        \Delta u_1+ \Delta u_2^* e^{ik} & -\Delta v
\end{pmatrix}
,
\end{equation}
where $\Delta v=\mathrm{Im}\,(v_1-v_2)/2$ and $\Delta u_n=-i(u_n^+- {u_n^-}^*)/2$.
This is a standard Hermitian dimer chain with two symmetric bands, encompassing the Su-Schrieffer Heeger \cite{Su79} and Rice-Mele \cite{Rice1982} models as special cases.
These models can support edge states, but they occur in the middle of the spectrum, while we are interested in the
upper edge of the spectrum. Furthermore, being Hermitian, $F$ does not suffer from  any complications of the skin effect, so that this upper edge can be found  from the Bloch version. Therefore, in large systems the causality threshold
\begin{equation}
\gamma_c=\sqrt{(\Delta v)^2+(|\Delta u_1|+|\Delta u_2|)^2}
\end{equation}
coincides for both types of boundary conditions, and hence is also insensitive to the skin effect.

To assess how visible these effects are in the passive setting, we use the spatially resolved density of states $\rho_n=Q_{nn}/2\pi$, again evaluated to the causality threshold.
Figure \ref{fig:NHSE} illustrates this in different scenarios \cite{Note1}. The top row shows results for a finite system with open boundary conditions,  where each panel further contrasts weak and strong coupling $\Gamma$.
In panel (a), the system supports an edge state, which for $\Gamma=0$ is clearly visible in the local and total density of states, while no such edge state exists in panel (b).
In both panels, the eigenstates display the non-Hermitian skin effect, as quantified by the summed  profile $I_n\equiv(UU^\dagger)_{nn}$ of normalized eigenvectors, collected in the diagonalizing matrix $U$.
However, this effect is completely disguised in the density of states, to the extent that it does not even appear as a background effect at finite $\Gamma$. Indeed, the spatial profile of the bulk density of states is similar to that in a system with periodic boundary conditions (bottom row), even though the eigenfunctions and eigenvalue spectra of both cases dramatically differ.

\emph{Conclusions.}
In summary, we have established general constraints on the observability of non-Hermitian effects in passive devices,
and evaluated the implications for prominent paradigms, in particular exceptional points, the non-Hermitian skin effect,
and symmetry-protected edge states. Some of the most widely studied features are effectively disguised in the density of states, in particular to the signatures of drastic mode nonorthogonality.
These findings highlight the essential role of active elements in devices that aim to exploit these signatures.

Being formulated in a unifying scattering approach, the general results apply to various platforms, such as photonic, mechanical and acoustic systems, electronic circuits, or microwave networks, and can be readily used to quantify the visibility of effects in a
wide range of models.
This also provides guidance for the design of passive devices in which the visibility is maximized, and allows to discard designs relying on effects whose observability is severely limited.

\begin{acknowledgements}
I gratefully thank Jan Wiersig for very useful discussions. All data produced in this work has been directly processed into the figures.
\end{acknowledgements}


%

\newpage

\appendix

\makeatletter
\renewcommand{\theequation}{S.\arabic{equation}}
\renewcommand{\thefigure}{S\@arabic\c@figure}
\makeatother
\setcounter{equation}{0}
\setcounter{figure}{0}

\section{Concrete forms of the symmetries of the nontrivial non-Hermitian content $F$}

In the main text, we identified the symmetries inherited by the operator $F$ in general terms. Here, we express this symmetries concretely in terms of the block structure of this operator.
We again start with the case of a PT-symmetric system. Taking $\mathcal{X}$ of the form of a Pauli-$x$ block matrix, as commonly done to reflect the physical design with symmetrically placed balanced gain and loss components, the effective Hamiltonian is of the form
\begin{equation}
H=\begin{pmatrix}
        A -i \gamma& B \\
        B^* & A^* -i \gamma
\end{pmatrix}
\end{equation}
with general subblocks $A$ and $B$.
The nontrivial non-Hermitian content is captured by the operator
\begin{equation}
F=\frac{-i}{2}\begin{pmatrix}
        A-A^\dagger & B-B^\dagger \\
        B^*-B^T & A^*-A^T
\end{pmatrix}.
\end{equation}
This displays the symmetries of a superconductor with a Hermitian charge-conjugation symmetry,
$\mathcal{X}F\mathcal{X}=-F^*$, and enforces the symmetry of its spectrum, with eigenvalues paired as $\pm |f_k|$.

For a PTT$'$ symmetry,
\begin{equation}
H=\begin{pmatrix}
        A -i \gamma& B \\
        C & A^\dagger -i \gamma
\end{pmatrix}
\end{equation}
where $B=B^\dagger$ and $C=C^\dagger$. Therefore,
\begin{equation}
F=\frac{-i}{2}\begin{pmatrix}
        A-A^\dagger & B-C \\
        C-B & A^\dagger-A
\end{pmatrix}
\end{equation}
now displays a chiral symmetry,  $\mathcal{X}F\mathcal{X}=-F$, again leading to a spectral symmetry of its eigenvalues.
The block structures  of the PT and PT$'$ variants coincide if the system is reciprocal, $H=H^T$, where $F$ is real and obeys time-reversal symmetry.

For systems with a C symmetry, one conventionally chooses $\mathcal{X}$ of the form of a Pauli-$z$ block matrix, as this allows for cases with a nontrivial topological index $\nu=\mathrm{tr}\,\mathcal{X}$. The effective Hamiltonian then has the structure
\begin{equation}
H=\begin{pmatrix}
        iA -i \gamma& B \\
        C & iD -i \gamma
\end{pmatrix}
\end{equation}
with real matrices $A$, $B$, $C$, $D$, and $\gamma$ chosen such that $\mathrm{tr} (A+D)=0$.
This entails that
\begin{equation}
F=\frac{-i}{2}\begin{pmatrix}
        i(A+A^T) & B-C^T \\
        C-B^T & i(D+D^T)
        \end{pmatrix}
        =\mathcal{X}F^*\mathcal{X}
\end{equation}
displays a generalized  time-reversal symmetry.

Analogously, for a system with CT$'$ symmetry,
\begin{equation}H=\begin{pmatrix}
        iA -i \gamma& B \\
        B^\dagger & iD -i \gamma
\end{pmatrix}
\end{equation}
 with general $B$ and Hermitian $A$ and $D$, once more obeying $\mathrm{tr} (A+D)=0$.
This entails that
\begin{equation}
F=\begin{pmatrix}
        A & 0\\
        0 & D
\end{pmatrix}
=\mathcal{X}F\mathcal{X}
\end{equation}
is indeed block-diagonalized into the symmetry sectors of $\mathcal{X}$.

Finally, in a non-reciprocal system with passive $T$ symmetry,
\begin{equation}
H=\begin{pmatrix}
        A-i \gamma& B \\
        C & D -i \gamma
\end{pmatrix}
\end{equation}
 with real blocks $A$, $B$, $C$, $D$, and
\begin{equation}
F=(-i/2)\begin{pmatrix}
        A-A^T& B-C^T \\
        C-B^T & D -D^T
\end{pmatrix}=-F^T,
\end{equation}
which indeed coincides with the block structure of the Hamiltonian for a topologically nontrivial superconductor in the Majorana basis \cite{Beenakker2015}.

\section{Analytical discussion of the visibility of exceptional points}
Here, we provide further analytical details for the signatures of mode nonorthogonality in the density of states near an exceptional point, as obtained from the model Hamiltonian Eq.~\eqref{eq:HEP}.
From the definitions,
this density of states can be written analytically as
\begin{align}
\rho(\omega)&
=\frac{2\mathrm{Re}\,[(a^2 + b c)(\gamma+\Gamma+i\omega)]}{\pi|a^2 + b c+(\gamma+\Gamma+i\omega)^2|^2}
\nonumber\\ &
+\frac{2\gamma|\gamma+\Gamma+i\omega|^2}{\pi|a^2 + b c+(\gamma+\Gamma+i\omega)^2|^2}
\nonumber\\ &
-\frac{\Gamma(2|a|^2+|b|^2+|c|^2)}{\pi|a^2 + b c+(\gamma+\Gamma+i\omega)^2|^2}
.
\end{align}
At the exceptional point $a^2+bc=0$, this reduces to
\begin{align}
\rho^{(\mathrm{EP})}(\omega)&=
\rho^{(1)}(\omega)+\rho^{(2)}(\omega),\\
\rho^{(1)}(\omega)&=\frac{2\gamma
}{\pi|\gamma+\Gamma+i\omega|^2},
\\
\rho^{(2)}(\omega)&=
-\frac{\Gamma(|b|+|c|)^2}{\pi|\gamma+\Gamma+i\omega|^4},
\end{align}
hence, the sum of a simple Lorentzian and a squared Lorentzian,
where the latter one only appears for finite $\Gamma$.
Equation~\eqref{eq:rhoep} is obtained for $\Gamma=0$,  $\gamma=\gamma_c=(|b|+|c|)/2$,
where the width $\gamma+\Gamma$ of the simple Lorentzian is minimized  in a passive system, and its weight
\begin{equation}
S_1(\Gamma)=\int_{-\infty}^\infty \rho^{(1)}(\omega)d\omega=\frac{2\gamma}{\gamma+\Gamma}
\end{equation}
is maximized.

The squared Lorentzian carries a negative weight
\begin{equation}
S_2(\Gamma)=\int_{-\infty}^\infty \rho^{(2)}(\omega)d\omega=-\frac{\Gamma(|b|+|c|)^2}{2(\gamma+\Gamma)^3}.
\end{equation}
Therefore, using again $\gamma\geq \gamma_c=(|b|+|c|)/2$, in a passive system the relative weight
\begin{equation}
\frac{|S_2(\Gamma)|}{S_1(\Gamma)}=\frac{\Gamma(|b|+|c|)^2}{4\gamma(\gamma+\Gamma)^2}\leq \frac{\Gamma\gamma}{(\gamma+\Gamma)^2}\leq \frac{1}{4},
\end{equation}
 where the maximum is attained at $\gamma=\Gamma=\gamma_c$.

Following the same steps, we can also compare the relative peak heights of these two contributions,
\begin{equation}
\frac{|\rho^{(2)}(0)|}{\rho^{(1)}(0)}=\frac{\Gamma(|b|+|c|)^2}{2\gamma(\gamma+\Gamma)^2}\leq \frac{1}{2},
\end{equation}
which again is maximized at $\gamma=\Gamma=\gamma_c$.
We note that the density of states formally turns negative for $\gamma<\gamma_c/\sqrt{2}$, which is less stringent than the causality constraint.

\section{Additional numerical results}

\begin{figure*}
\includegraphics[width=\linewidth]{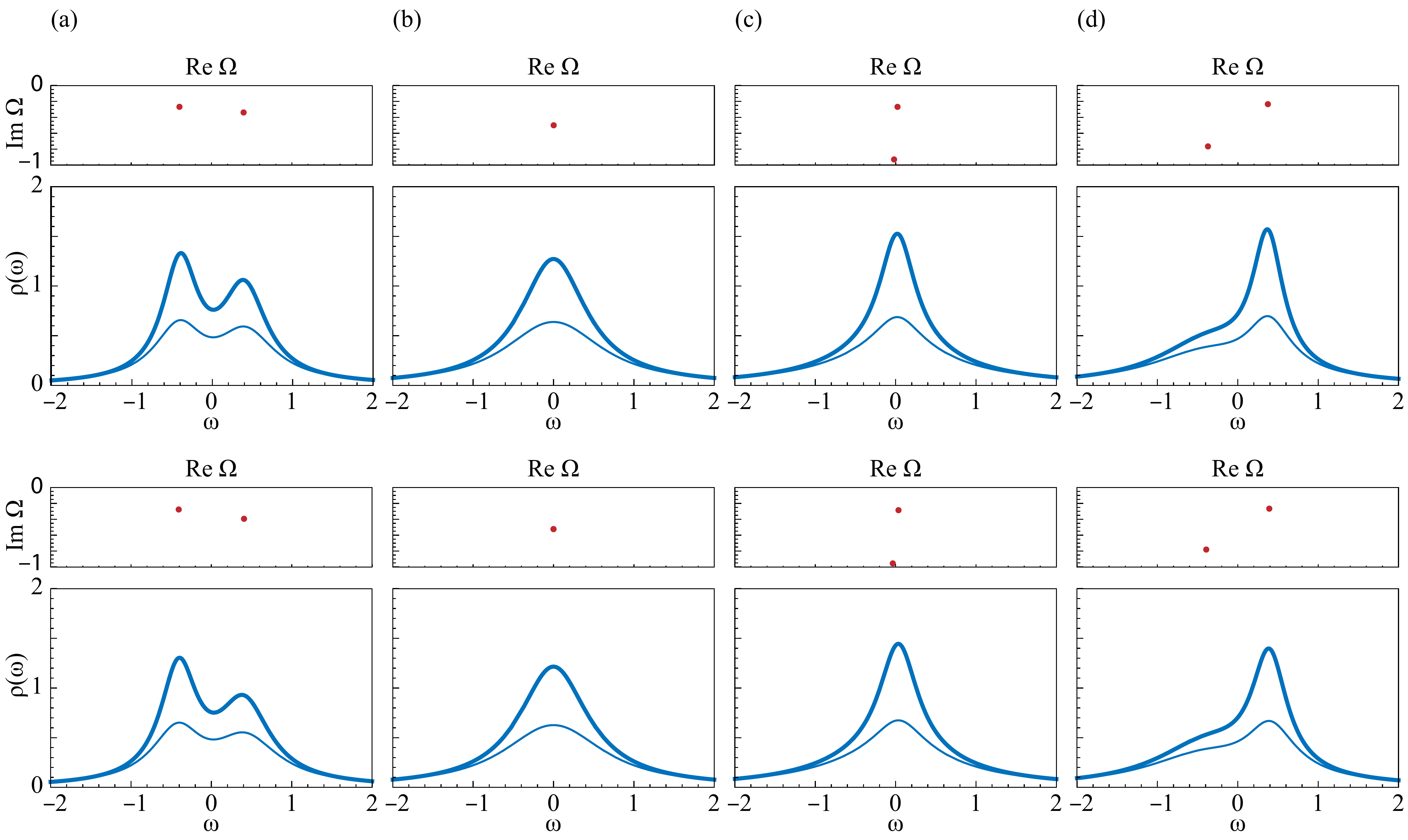}
 \caption{Analogous to Fig.~\ref{fig:EP} in the main text, but with complex non-reciprocal coupling parameters
$b = 0.3 - 0.4 i$,
$c = 0.4 + 0.3 i$ (top) $b = 0.6$,
$c = 0.4 - 0.2 i$, while again (a) $a=\sqrt{-bc}-0.2 i$,
  (b) $a=\sqrt{-bc}$ (EP),  (c) $a=\sqrt{-bc}+0.1 i$, and (d) $a=\sqrt{-bc}+0.2$.  At the exceptional point, $\rho(\omega)$ is again a simple Lorentzian.
  }\label{fig:App1}
\end{figure*}

\begin{figure*}
  \includegraphics[width=\linewidth]{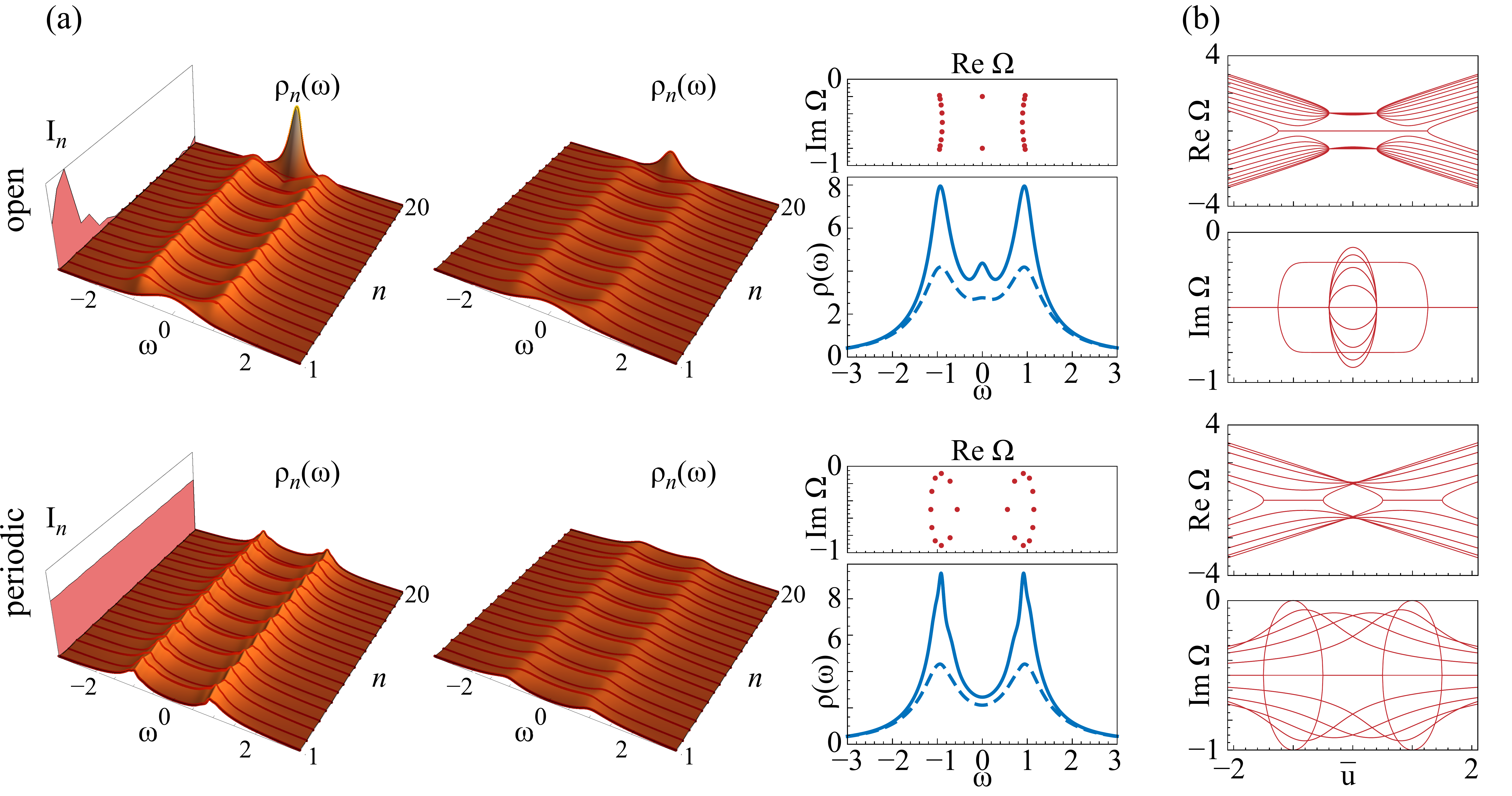}
 \caption{(a) Analogous to Fig.~\ref{fig:NHSE} in the main text, but for $\bar u=0.25$, where the bulk modes in the system with open boundary conditions have moved away from the symmetry line in the complex frequency plane.
$(b)$ Evolution of the eigenvalues with the parameter $\bar u$.
  }\label{fig:App2}
\end{figure*}

\begin{figure*}
  \includegraphics[width=\linewidth]{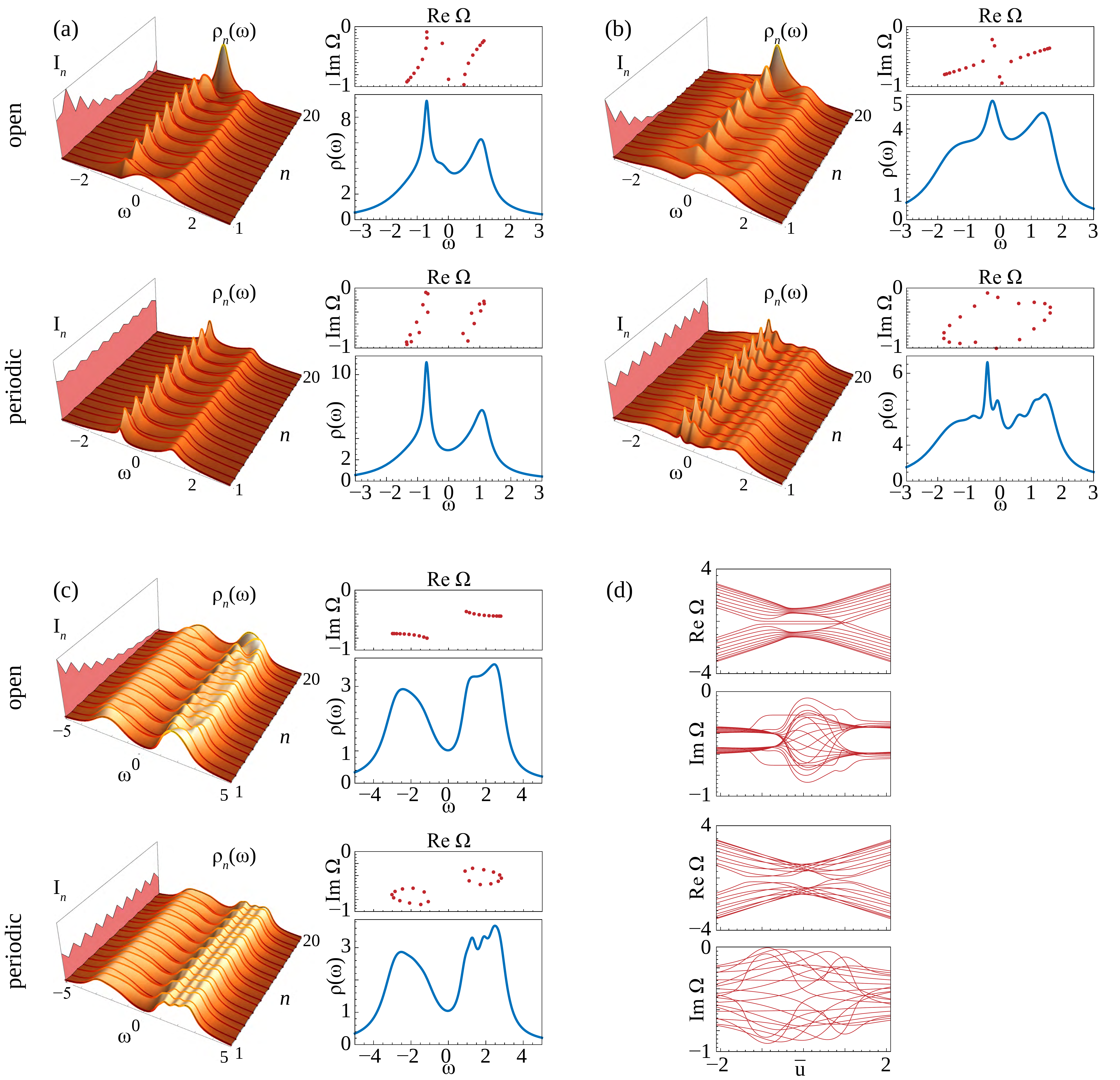}
 \caption{Analogous to Figs.~\ref{fig:NHSE} and \ref{fig:App2}, but for $\Gamma=0$ only, and the other parameters set to
 $v_1=-0.3i-i \gamma_c $, $v_2=-0.2+0.3i-i \gamma_c$, $u_1^+=\bar u- 0.4+0.4i$,
   $u_1^-=\bar u+ 0.4-0.1i$, $u_2^+=1-0.1i$, $u_2^-=0.9$,
  $w_1=w_2=0$, and  (a) $\bar u=0.25$, (b) $\bar u=0.8$, (c) $\bar u=2$. Panel (d) shows the evolution of the eigenvalues with the parameter $\bar u$.}\label{fig:App3}
\end{figure*}

\begin{figure*}
  \includegraphics[width=\linewidth]{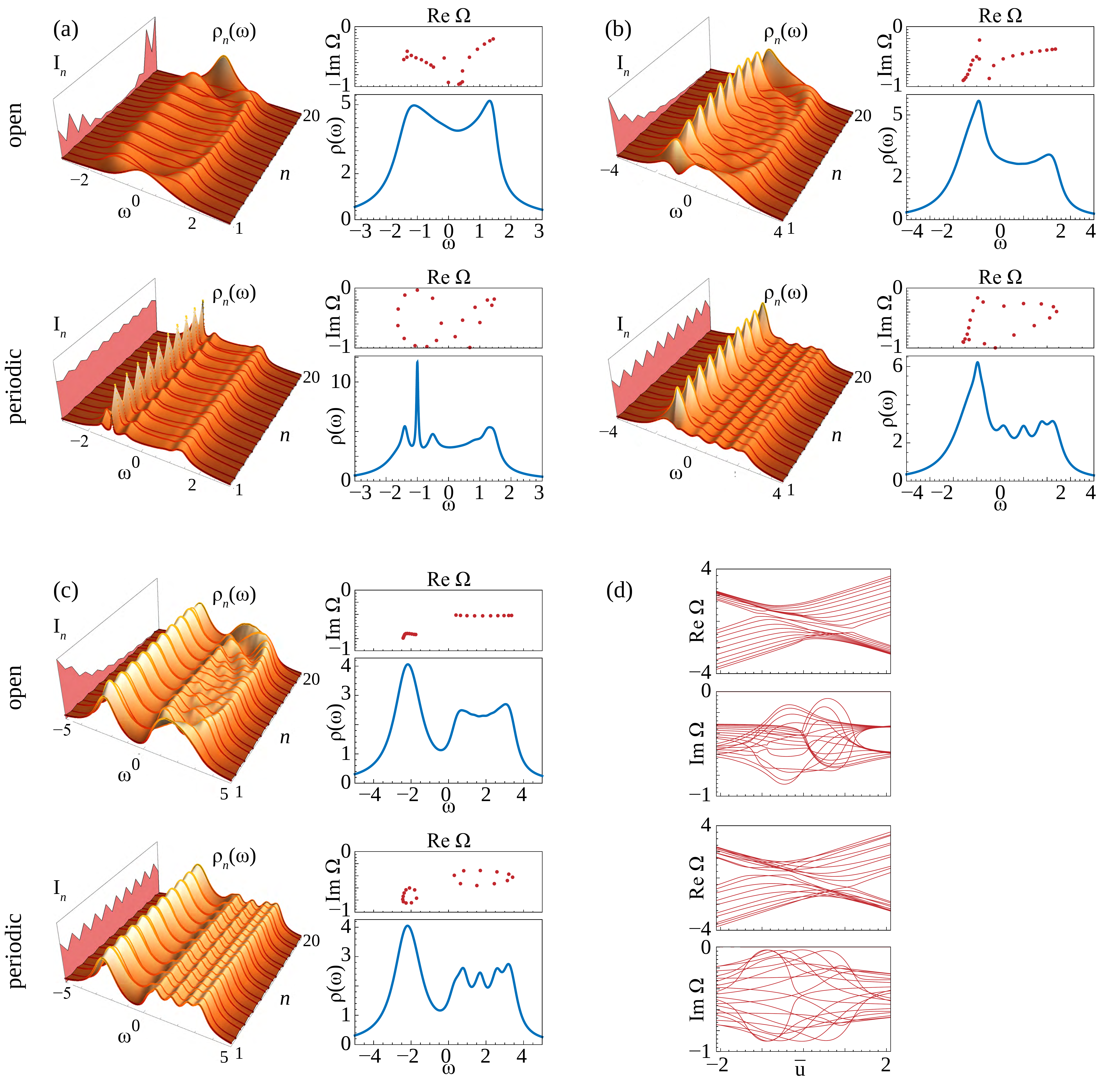}
 \caption{Analogous to Fig.~\ref{fig:App3}, but including next-nearest neighbour couplings
   $w_1=0.2$, $w_2=0.4$, while here  (a) $\bar u=0$, (b) $\bar u=1$, (c) $\bar u=2$. }\label{fig:App4}
\end{figure*}

In the main text we showed numerical results for particularly interesting scenarios in which the underlying non-Hermitian effects are realized cleanly. To illustrate the general nature of our findings, we show in Fig.~\ref{fig:App1} additional results for the EP model \eqref{eq:HEP}, evaluated at parameters where PT and PTT$'$ symmetries are manifestly broken also by the couplings.
Analogously, we show additional results for the non-Hermitian skin effect, covering the case where some bulk states in the open system have moved away from the symmetry line in the complex plane (Fig.~\ref{fig:App2}), as well as parameter configurations in which the
all spectral symmetries are explicitly broken (Fig.~\ref{fig:App3}), including by next-nearest-neighbour couplings  (Fig.~\ref{fig:App4}).

\end{document}